# Dual nature of magnetic dopants and competing trends in topological insulators

**Paolo Sessi[1,§], Rudro R. Biswas[2], Thomas Bathon[1], Oliver Storz[1], Stefan Wilfert[1], Alessandro Barla[3], Konstantin A. Kokh[4,5,6], Oleg E. Tereshchenko[5,6,7], Kai Fauth[1,8], Matthias Bode[1,8], Alexander V. Balatsky[9,10]**

[1] Physikalisches Institut, Experimentelle Physik II, Universität Würzburg, Am Hubland, 97074 Würzburg, Germany

[2] Department of Physics and Astronomy, Purdue University, 525 Northwestern Ave, West Lafayette, Indiana 47907, USA

[3] Istituto di Struttura della Materia (ISM), Consiglio Nazionale delle Ricerche (CNR), 34149 Trieste, Italy.

[4] V.S. Sobolev Institute of Geology and Mineralogy, Siberian Branch, Russian Academy of Sciences, 630090 Novosibirsk, Russia

[5] Novosibirsk State University, 630090 Novosibirsk, Russia

[6] Saint-Petersburg State University, 198504 Saint-Petersburg, Russia

[7] A.V. Rzanov Institute of Semiconductor Physics, Siberian Branch, Russian Academy of Sciences, 630090 Novosibirsk, Russia

[8] Wilhelm Conrad Röntgen-Center for Complex Material Systems (RCCM), Universität Würzburg, Am Hubland, 97074 Würzburg, Germany

[9] Institute for Materials Science, Los Alamos National Laboratory, Los Alamos, New Mexico 87545, USA

[10] Nordita, Center for Quantum Materials, KTH Royal Institute of Technology, and Stockholm University, Roslagstullsbacken 23, 106 91 Stockholm, Sweden

§ To whom correspondence should be addressed.
Email: sessi@physik.uni-wuerzburg.de

**Abstract:**

**Topological insulators interacting with magnetic impurities have been reported to host several unconventional effects. These phenomena are described within the framework of gapping Dirac quasiparticles due to broken time-reversal symmetry. However, the overwhelming majority of studies demonstrate the presence of a finite density of states near the Dirac point even once Topological insulators become magnetic. Here, we map the response of topological states to magnetic impurities at the atomic scale. We demonstrate that magnetic order and gapless states can coexist. We show how this is the result of the delicate balance between two opposite trends, i.e. gap opening and emergence of a Dirac node impurity band, both induced by the magnetic dopants. Our results evidence a more intricate and rich scenario with respect to the once generally assumed, showing how different electronic and magnetic states may be generated and controlled in this fascinating class of materials.**

**Introduction**

Topological insulators (TIs) represent a new state of matter where electronic properties are not related to any symmetry breaking but dictated by the concept of topology. Beyond their fundamental interest, the separation of electronic states into a gapped bulk and helical spin Dirac cone at the surface offers great potential for applications. Ultimately, the success of TIs crucially depends on the ability of functionalizing them with well-defined perturbations. Within this framework, magnetic dopants occupy a primary role [1-18]. Since they break the time-reversal symmetry which protects topological states, they are expected to modify the Dirac spectrum while at the same time changing the spin-texture [5-8]. This interaction is expected to result in several exotic phenomena with interesting implications in spintronics and magneto-electric [1-4]. The recent experimental realization of the quantum anomalous Hall effect confirmed the presence of a gapped Dirac state [3,4]. On the other hand, the majority of photoemission [9-13] and local tunneling data [14-18] clearly demonstrate the presence of a finite density of states (DOS) near the Dirac node in magnetically doped TIs. The disorder induced by the dopants further complicates the story since it generates spatial fluctuations of the chemical potential [16] which significantly broaden the lineshape of photoemission spectra potentially masking the detection of a gap [9-13]. Recent theoretical predictions suggested that the resolution of this apparent paradox might be directly linked to the dual nature of the electronic states generated by magnetic impurities [19]. However, these claims escaped experimental verification.

In the following, we provide compelling evidence that on one hand the emergence of ferromagnetism gaps the Dirac states; on the other hand impurity-induced resonances form an impurity band at or near the Dirac node filling the gap. Both effects originate from the same magnetic impurity doping, leading to the emergence of a two-fluid behavior. Our results resolve long-standing contradictions in the field. They paint a more complicated and multifaceted story of magnetically doped TIs and provide a detailed microscopic picture which allows engineering electronic and magnetic properties by acting on the delicate balance in between the impurity band and the gapped Dirac states.

## Results

### Structural and magnetic properties

Fig. 1a shows a photographic image representative of the bulk single crystals used in the present study, i.e. vanadium (V)-doped $Sb_2Te_3$. This system recently led to the robust experimental realization of the quantum anomalous Hall effect [4], thus being an optimal platform to investigate the interaction of topological states with time reversal symmetry breaking perturbation. The nominal stoichiometry corresponds to $Sb_{1.985}V_{0.015}Te_3$. Consequently, within the typical quintuple-layered structure characterizing these materials, V atoms substitute Sb as sketched in Fig. 1b at a concentration of 0.75 % in each Sb plane. Fig. 1c shows a typical large scale topographic image of the samples acquired by scanning tunneling microscopy (STM) in the constant-current mode. Dark points correspond to V atoms. Their introduction results in the creation of local perturbations in the crystal structure, which are imaged as depressions at the surface. Note that the V atoms are homogeneously distributed over the surface, without any clustering. This observation proves the good quality of the crystals used in our experiments, which can be thus classified as diluted doped. A quantitative estimation of the V concentration can be achieved by counting the number of depression visible on the surface, giving a value of 0.725 % in each Sb layer, in good agreement with the nominal concentration of the crystals. Fig. 1d shows an atomically resolved image of the area highlighted by a black box in Fig. 1c. This zoomed image allows visualizing two different V species marked by arrows. An analysis of their appearance based on their symmetry and surface extension allows, as described in Ref. [20], to identify them as V atoms located in the 2$^{nd}$ (A) and 4$^{th}$ Sb (B) atomic layer, respectively.

To investigate the magnetic response of the topological insulator to the introduction of magnetic dopants, x-ray magnetic circular dichroism (XMCD) measurements have been performed at the PM3 bending magnet beamline of HZB [21] with a custom XMCD end station. X-ray absorption was recorded in the total electron yield (TEY) mode in the presence of an external magnetic field, by measuring the photoinduced sample drain current, as schematically illustrated in Fig. 1e. The short probing depth of this technique [22] allows investigating the existence of magnetic order at the surface, i.e. where the topologically protected states exist. Fig. 1f reports the x-ray absorption spectra taken at the V $L_{2,3}$ edges at a temperature of $T = 13.2$ K in a magnetic field of $\pm 1$ T applied along the surface normal. Their difference (=XMCD) (green line) highlights the considerable magnetic polarization of the near-surface V dopants and hence the existence of uncompensated magnetic moments. An XMCD hysteresis curve acquired at the same temperature is reported in Fig. 1g. Its square shape proves the ferromagnetic nature of this polarization and a high degree of homogeneity (absence of domain wall pinning) while the full remanence characterizes the surface normal as the easy direction of magnetic anisotropy.

### Impact of impurities on topological states and emergence of Dirac node excitations

To visualize the impact of impurity ferromagnetism on the topological states, scanning tunneling spectroscopy (STS) measurements have been performed on the very same crystals at $T = 4.8$ K, i.e. well below the Curie temperature. The local density of states (LDOS) for energies centered around the Dirac point, i.e. where a gap is expected to open, is reported in Fig. 2a by positioning the tip away from any defects. The steep increase of the conductance visible at approximately

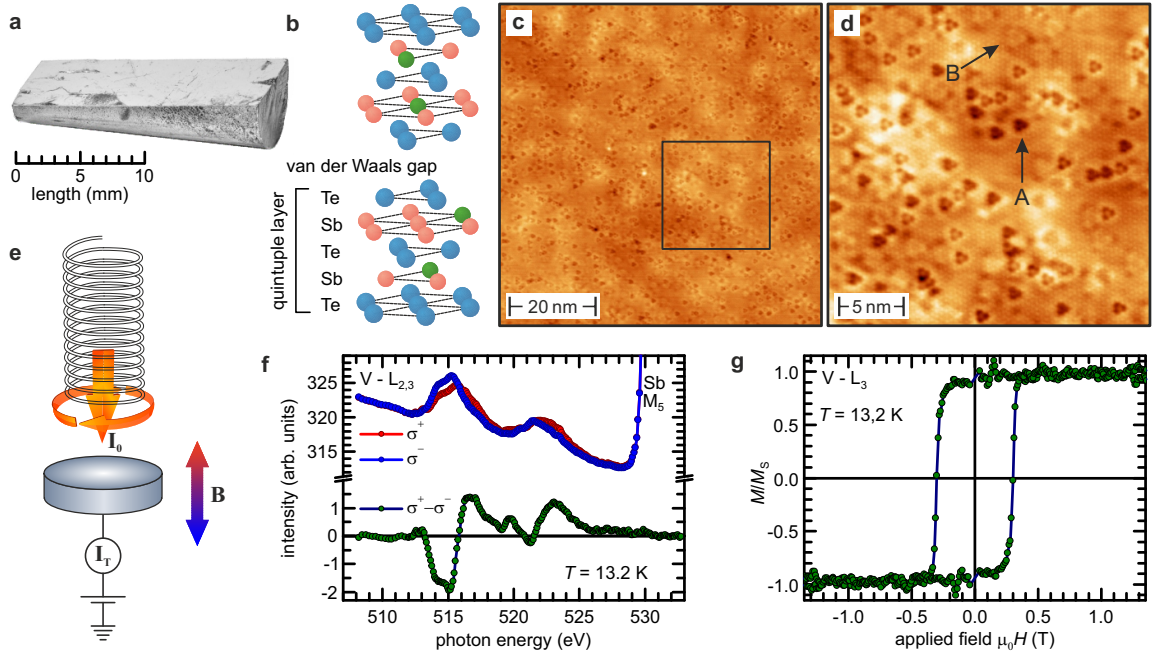

**Figure1: Structural and magnetic properties of V-doped $Sb_2Te_3$**. (**a**) Vanadium bulk-doped $Sb_2Te_3$ single crystal ($Sb_{1.985}V_{0.015}Te_3$); (**b**) $Sb_2Te_3$ crystal structure: V atoms (green) are located in the $2^{nd}$ and $4^{th}$ layer, substituting Sb; (**c**) 75×75 nm$^2$ topographic image showing the absence of any clustering; (**d**) high resolution image showing the V substitutional sites. Triangular depressions labelled as A and B correspond to the two possible V substitutional sites located in the $2^{nd}$ and $4^{th}$ layer, respectively; (**e**) schematic of XMCD measurements. Photons with well-defined circular polarization and energy are absorbed by the sample in a variable magnetic field. The resulting photocurrent (TEY) is measured, making the technique sensitive to surface magnetism. (**f**) TEY signals for opposite magnetic fields (blue and red lines) of $B = \pm 1$ T. Their difference (green line) sheds light on the magnetic response of the V dopants; (**g**) out-of-plane V-$L_3$ XMCD magnetization curve: the square shape of the hysteresis loop proves the existence of long-range surface ferromagnetic order.

50 meV and 200 meV above the Fermi level corresponds to the onset of the valence and conduction band, respectively. Within this energy range, the conductance is strongly suppressed because of the bulk gap [20]. Spectroscopic measurements optimized over this energy range reveal that, although strongly reduced, the conductance never goes to zero but simply reaches a minimum at approximately 175 meV. Unlike in the pristine scenario, where a pure V-shape characterizes STS spectra inside the bulk-gap [20,23], this minimum is energetically close to but not directly at the Dirac point (see below). These data unequivocally demonstrate that, contrary to what is generally assumed, a spectral gap does not open at the Dirac point despite the presence of long-range surface magnetic order which breaks the time reversal symmetry protecting topological states.

To shed light on the origin of the finite LDOS, STS measurements have been performed by positioning the tip on top of the V dopants. This allows to directly investigate the electronic response that, in addition to the magnetic one highlighted by the XMCD measurements, they introduce into the system. In both cases, i.e. V atoms occupying the $2^{nd}$ and $4^{th}$ layer (Fig. 2b and c, respectively), STS spectra visualize the emergence of strong resonances energetically located close to the Dirac point [24]. Fig. 2d and e map the spatial distribution of their intensity, revealing how their spectral weight is not significantly suppressed over distances of few nanometers, i.e. the same length scale as the average distance between dopants (see Fig. 1c). These impurity-induced resonances are thus significantly overlapping and likely form an impurity band centered at the resonance energy with a non-negligible spectral weight over the entire surface. Note that the energy position of these resonances is always in close proximity of the Dirac point, i.e. where a gap is expected to open due to the magnetically ordered state.

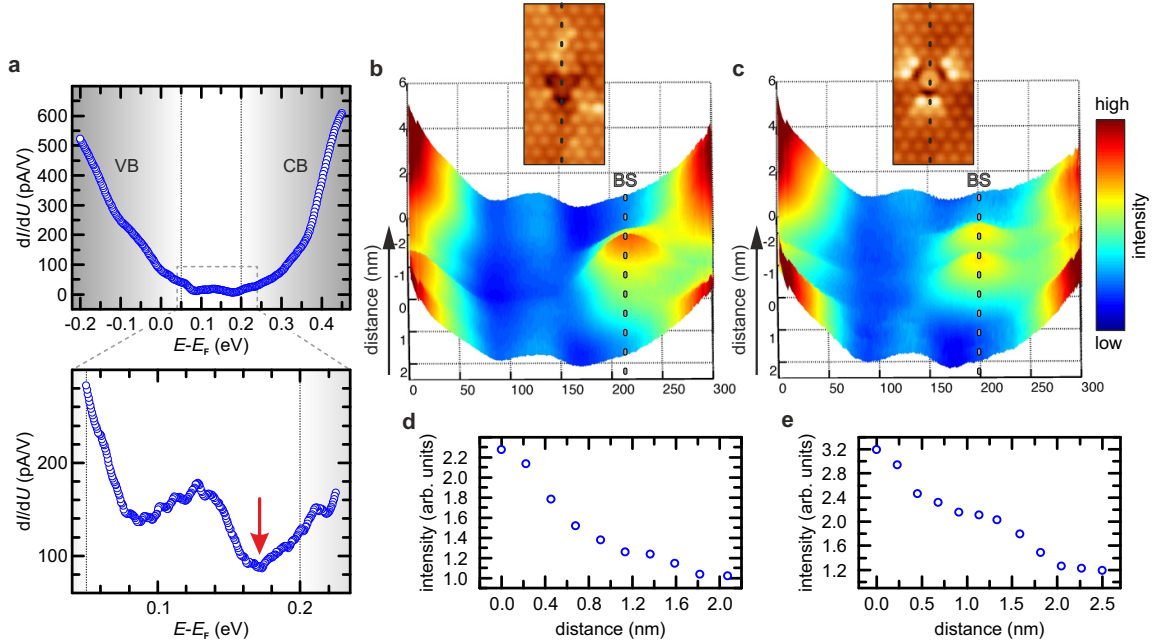

**Figure 2 Spectroscopic mapping of impurity-induced states**. (**a**) Scanning tunneling spectroscopy of the ferromagnetic $Sb_{1.985}V_{0.015}Te_3$. Data have been acquired by positioning the tip away from any defect. The density of states reaches a minimum (see arrow) at approximately 175 meV above the Fermi level. The spectra reveal the absence of any gap opening despite the existence of long-range ferromagnetic order (see zoomed area) Stabilization parameters: $I = 50\,\text{pA}$, $V = 450\,\text{mV}$, zoomed area $I = 50\,\text{pA}$, $V = 225\,\text{mV}$; (**b,c**) spectroscopic measurements taken along lines crossing the two different V sites (see insets). In both cases strong resonances energetically located close to the Dirac point (indicated by BS) emerge in the nearby of the dopants; (**d,e**) spatial dependence of the dopants induced resonances for **b** and **c**, respectively. The zero in the distance axis corresponds to the position of maximum intensity. In both cases the spectral weight decays over distances of few nanometers. $T = 4.8$ K.

In the following, we demonstrate that their experimental detection is consistent with the expected universal response to perturbations of Dirac materials [25], i.e. materials such as d-wave superconductors, graphene and, like in the present case, topological insulators, whose low energy excitation spectra is described by massless Dirac particles rather than the Schrödinger equation. This implies that the behavior of topological states with respect to time-reversal symmetry breaking cannot always be correctly described within the generally assumed framework of gapless vs. magnetically gapped states, but that it reflects the competition between the trends for gapping due to magnetic scattering and gap filling due to impurity resonances filling up the gap [6,7,19].

**Dual nature of magnetic dopants: theory discussion**

Once local impurities with both magnetic ($J\mathbf{S}$) and scalar potential ($U$) scattering components are introduced in the system [6-7,24-25], the Hamiltonian can be expressed as:

$$\mathcal{H} = (\mathbf{k}_x \boldsymbol{\sigma}_x + \mathbf{k}_y \boldsymbol{\sigma}_y) + \Sigma_{\text{imp}}[J\mathbf{S}_i \cdot \boldsymbol{\sigma}(\mathbf{r}_i) + Un(\mathbf{r}_i)] \quad . \tag{1}$$

Here, $J$ and $U$ are the impurity magnetic exchange and potential strengths, respectively, while $\mathbf{k}$ is the wave vector. $n(\mathbf{r}_i)$ and $\boldsymbol{\sigma}(\mathbf{r}_i)$ are the electronic number and spin densities, respectively, at the location, $\mathbf{r}_i$, of the i$^{\text{th}}$ impurity. We have chosen units such that the Planck's constant, $\hbar$, and the Fermi (Dirac) velocity, $v_{\text{D}}$, are both unity.

When the impurities align ferromagnetically perpendicular to the surface (z-direction), it is usually assumed that an energy gap appears at the Dirac point as a result of the spatially averaged carrier-impurity exchange interaction. However, this analysis ignores the role of the impurity potential strength, $U$, which leads to intragap resonances [6,7,19]. Taking this contribution into account, the effective Hamiltonian, including a mean-field gap arising from the magnetic moments, becomes

$$\mathcal{H}_{\text{eff}} = (\mathbf{k}_x \boldsymbol{\sigma}_x + \mathbf{k}_y \boldsymbol{\sigma}_y + \Delta \boldsymbol{\sigma}_z) + \Sigma_{\text{imp}}[J_{\text{eff}} \mathbf{S} \cdot \boldsymbol{\sigma}_z + U] n(\mathbf{r}_i) \quad . \tag{2}$$

The mass,

$$\Delta \simeq J \Sigma_{\text{imp}} \langle n(\mathbf{r}_i)(\mathbf{S}_i)_z \rangle \quad , \tag{3}$$

is half the energy gap of the now massive surface Dirac states, found from an averaging procedure over the impurity-carrier exchange interactions, and $J_{\text{eff}}$ is the residual effective magnetic impurity strength of the impurity acting on these *massive* Dirac fermions. Calculating the exact Green's function of this effective Hamiltonian with the T-matrix method [6] in the presence of a single impurity, we find analytically that when the scalar potential strength, $U$, is larger in magnitude than the residual exchange, $J_{\text{eff}}\mathbf{S}$, bound states appear inside the magnetization-induced energy gap. The bound states appear and fill up the spectral gap as shown in Fig. 3. This finding demonstrates that our theoretical approach correctly captures the experimental data (Figs. 2a-e). For a more detailed discussion, see Supplementary Note 1.

These results paint a more intricate and rich scenario than the general assumption of a spectral gap appearing in magnetic TIs. In particular, they demonstrate the emergence of a two-fluid behavior induced by the magnetic dopants which drive a competition in between two different and opposite trends: gap opening vs. gap filling which, as in the present case, can coexist.

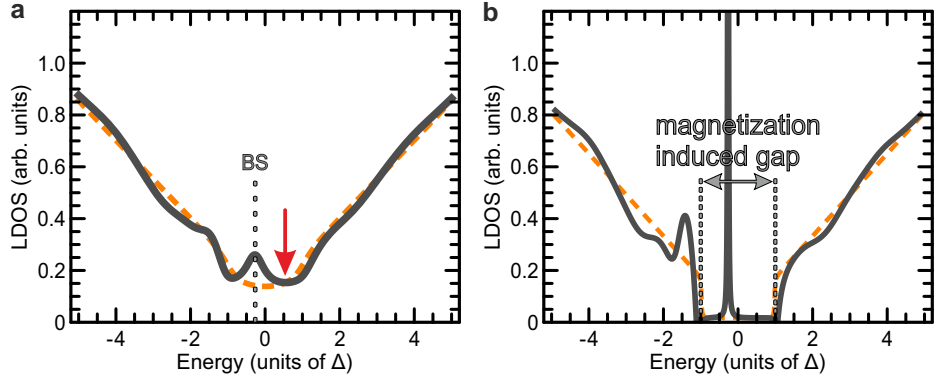

**Figure 3 Dual-response of topological states to magnetic impurities**. (**a**) theoretically calculated LDOS at a location with many nearby impurities, using the mean field theory, Eq. (2). In units where $\hbar$, the Fermi velocity and the gap parameter, $\Delta$, are unity, we have used the impurity strength parameters $J_{\text{eff}}S = 0.5$, $U = 1.75$, high energy cutoff $\Lambda = 50$, and the energy broadening parameter $= 0.3$. Impurities are located at distances $r = 0.6, 1.4, 1.5, 1.6, 1.8, 1$ and 3 to mimic their experimental distribution. We have ignored multi-impurity scattering, which would lead to further broadening of the bound state feature (indicated as BS). The dashed curve is the LDOS in the absence of local impurity perturbation. As illustrated in **b** for negligible energy broadening ($= 0.01$), a clear magnetization-induced energy gap exists between $\pm\Delta$. When the absolute value of $U(J_{\text{eff}}S)^{-1}$ is larger than a threshold (equal to 1 for $\Gamma = 0$), bound states appear inside the gap. Thus, experimental measurements of the LDOS may never show a spectral gap, even though the mobility (transport) gap is nonzero because the localized intra-gap bound states do not affect long range transport at low temperatures.

**Two-fluid behavior and mobility gap**

This behavior might sound counterintuitive or even in conflict with the observation of phenomena such as the quantum anomalous Hall effect within this class of systems. In the following we demonstrate that this is not the case, and that this apparent discrepancy reflects the localized vs. delocalized nature of the impurity induced states, and their different response to electric and magnetic fields. This can be directly visualized by performing Landau level (LL) spectroscopy at high magnetic fields. Fig. 4a-d reports the results obtained on pristine $Sb_2Te_3$. For sufficiently strong magnetic fields applied perpendicular to the surface, a sequence of peaks starts to emerge from the background, signaling the condensation of the 2D topological states into a well-defined sequence of LLs [26]. As expected for linearly dispersing states [27], all LLs can be nicely fitted by the relation:

$$E_n - E_D = \text{sgn}(n)\hbar v_D \sqrt{\frac{2e|n|B}{\hbar}}$$

(4)

where $n$ is the Landau level index, $B$ the magnetic field, and $v_D$ the Dirac velocity which describes the LLs energy sequence for massless particles.

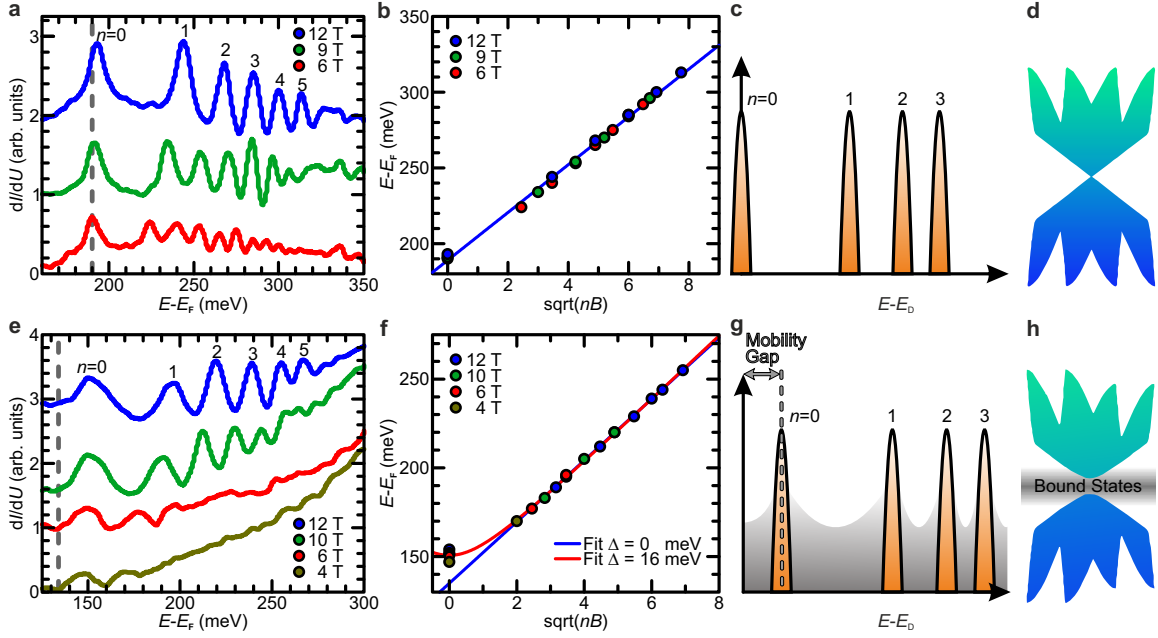

**Figure 4 Emergence of a two-fluid behavior**. Landau levels measured by scanning tunneling spectroscopy in high magnetic fields on a pristine (**a-d**) and (**e-h**) V-doped $Sb_2Te_3$. The spectrum obtained at zero-field has been subtracted in both cases. Spectra at different magnetic fields have been vertically shifted for clarity. The number $n$ identifies the Landau level index. While all Landau levels of pristine $Sb_2Te_3$ can nicely be fit linearly, this is not the case for $Sb_{1.985}V_{0.015}Te_3$. In this case, the onset of ferromagnetism shifts the zero Landau level towards higher energies, signaling the mass acquisition of the helical Dirac fermions. Panel **g** illustrates the creation of a mobility gap despite the existence of a finite density of states generated by the magnetic impurities (gray shadow). As schematically illustrated in **h**, these states are filling the gap opened by the broken time-reversal symmetry. Their existence is intimately linked to the dual nature of the dopants, which result in the emergence of a two-fluid behavior with opposite and competing trends.

Remarkably, once the same measurements are performed on the magnetically ordered V-doped samples (see Fig. 4e-h), the zero order LL does not follow the fit, but shifts towards higher energies. This is highlighted by the substantial shift of the first peak visible in Fig. 4e with respect to the dashed gray line which marks the position of the zero order LL as expected for linearly dispersing Dirac states (raw STS data before zero-field background subtraction are reported in supplementary figure 3). This behavior unambiguously signals the emergence of a mobility edge for the topological states. The emergence of mobility edges is well-documented in two-dimensional conventional quantum Hall states [28]. In TIs, the evolution of the Landau levels for Dirac states, mixed with a continuum of localized states, has the dispersion

$$E_n - E_D = \begin{cases} sgn(n)\sqrt{2eB\hbar v_D^2|n| + \Delta^2} &, n \neq 0 \\ \Delta &, n = 0 \end{cases}$$

(5)

Here, with respect to the pristine case, the appearance of $\Delta$ signals the acquisition of a mass term for the Dirac states. As shown in Fig. 4f, this equation describes the experimental data very well, the main effect of the mass term being manifested in a shift of the zero-order Landau level. Concurrently, the magnetic dopants also produce a continuum of localized states as indicated by the shaded regions in Fig. 4g,h. These states also fill the spectral gap about the Dirac point. In agreement with our theoretical arguments, we have found that these are mainly localized states bound to magnetic dopants (see Fig. 2d,e). Consequently, once the chemical potential is tuned to their energy range, a mobility gap appears, enabling a full anomalous quantum Hall effect quantization while at the same time inducing a finite density of states at all energies.

**Discussion**

In conclusion, we demonstrated that magnetic dopants induce multiple changes with opposite trends in the excitation spectra of TIs. Contrary to what generally believed, this competition may give rise to the coexistence of long-range magnetic order and gapless states. The microscopic origin of this behavior is directly linked to the dual nature of the dopants which on one hand induce a mass term in the Dirac spectrum while at the same time create intragap states with large spectral density *within the magnetic gap*. Overall, this delicate balance results in the emergence of a two fluid behavior where the localization of the impurity resonances produces a mobility gap. Our observations call for a reevaluation of the claims about magnetically doped TI. More generally, they provide evidence of the richness of the physics of TIs demonstrating how, going beyond a single particle approach, it is possible to tailor the competition between opposite trends paving the way to engineer novel spintronics and magneto-electric effects in this fascinating class of materials.

**Acknowledgments**

Work was supported by US DOE E304, VR and KAW (A.V.B.) and by DFG through SFB 1170 "ToCoTronics", project A02 (P.S., T.B., O.S., S.W., M.B.). R.R.B. thanks Purdue University Startup funds for support. Allocation of synchrotron radiation beamtime as well as financial support by HZB is gratefully acknowledged. We are grateful to A. Black-Schaffer, J. Fransson and T. Wehling for useful discussions.

**Authors Contribution**

P.S., T.B., O.S. and S.W. performed and analyzed STM and STS measurements. T.B., A.B. and K.F. performed XMCD measurements, K.F. analyzed them. K.A.K. and O.E.T. synthesized the crystals. R.R.B. and A.V.B. provided the theoretical framework. All authors contributed writing the manuscript.

**Methods**

**Crystal growth:** Single crystals of about 10 mm in diameter and 60 mm in length were grown by the modified vertical Bridgman method with rotating heat field [29]. Required stoichiometric amounts of Sb, V and Te were loaded to a carbon-coated quartz ampoule. After evacuation to $p = 10^{-4}$ Torr the ampoule translation rate and axial temperature gradient were set to 10 mm day$^{-1}$ and 15 °C cm$^{-1}$, respectively.

**STM measurements:** Single crystals have been cleaved at room temperature in ultra-high-vacuum and immediately inserted into the STM operated at cryogenic temperatures. All measurements have been performed using electrochemically etched tungsten tips. Spectroscopic data have been obtained using the lock-in technique by adding a bias voltage modulation in between 1 and 10 meV (r.m.s.) at a frequency of $f = 793$ Hz. Data displayed in Figure 1 and 2 have been obtained using a commercial STM operated at $T = 4.8$ K. Landau level spectroscopy data (Figure 4) have been obtained using a home-built STM operated at $T = 1.3$ K. The exact energy position of the Landau levels has been obtained by considering the maximum of a gaussian function fitting each peak after background subtraction.

**Data availability:** The authors declare that the data supporting the findings of this study are available within the article and its supplementary information. Data files are available from the corresponding author upon request.

**Competing financial interests**

The authors declare no competing financial interests.

**References:**

[1] Qi, X-.L., Li, R., Zang, J. & Zhang S.-C. Inducing a magnetic monopole with topological surface states. *Science* **232**, 1184-1187 (2009).

[2] Garate, I. & Franz M. Inverse spin-galvanic effect in the interface between a topological insulator and a ferromagnet. *Phys. Rev. Lett.* **104**, 146802 (2010).

[3] Chang, C.-Z. *et al.* Experimental Observation of the quantum anomalous hall effect in a magnetic topological insulator. *Science* **340**, 167-170 (2013).

[4] Chang, C.-Z. *et al.* High-precision realization of robust quantum anomalous hall state in a hard ferromagnetic topological insulator. *Nat. Mat.* **14**, 473-477 (2015).

[5] Liu, Q., Liu, C.-X., Xu, C., Qi, X.-L. & Zhang, S.-C. Magnetic impurities on the surface of a topological insulator. *Phys. Rev. Lett.* **102**, 156603 (2009).

[6] Biswas, R. R. & Balatsky, A. V. Impurity-induced states on the surface of three-dimensional topological insulators. *Phys. Rev. B* **81**, 233405 (2010).

[7] Black-Schaffer, A. M. & Balatsky, A. V. Strong potential impurities on the surface of a topological insulator. *Phys. Rev. B* **85**, 121103(R) (2012).

[8] Zhu, J.-J., Yao, D.-X., Zhang, S.-C. & Chang, K. Electrically controllable surface magnetism on the surface of topological insulators. *Phys. Rev. Lett.* **106**, 097201 (2011).

[9] Chen, Y. L. *et al.* Massive dirac fermion on the surface of a magnetically doped topological insulator. *Science* **329**, 659-662 (2010).

[10] Xu, S.-Y. *et al.* Hedgehog spin texture and berry's phase tuning in a magnetic topological insulator. *Nat. Phys.* **8**, 616-622 (2012).

[11] Valla, T., Pan, Z.-H., Gardner, D., Lee, Y. S., Chu, S. Photoemission spectroscopy of magnetic and nonmagnetic impurities on the surface of the Bi2Se3 topological insulator. *Phys. Rev. Lett.* **108**, 117601 (2012).

[12] Scholz, M. R. *et al.* Tolerance of topological states towards magnetic moments: Fe on Bi2Se3. *Phys. Rev. Lett.* **108**, 256810 (2012).

[13] Schlenk, T. *et al.* Controllable magnetic doping of the surface state of a topological insulator. *Phys. Rev. Lett.* **110**, 126804 (2013).

[14] Hor, Y. S. *et al.* Development of ferromagnetism in the doped topological insulator $Bi_{2-x}Mn_xTe_3$. *Phys. Rev. B* **81**, 195203 (2010).

[15] Okada, Y. *et al.* Direct observation of broken time-reversal symmetry on the surface of a magnetically doped topological insulator. *Phys. Rev. Lett.* **106**, 206805 (2011).

[16] Beidenkopf, H. *et al.* Spatial fluctuations of helical dirac fermions on the surface of topological insulators. *Nat. Phys.* **7**, 939-943 (2011).

[17] Yang, F. *et al.* Identifying magnetic anisotropy of the topological surface state of $Cr_{0.05}Sb_{1.95}Te_3$ with spin polarized STM. *Phys. Rev. Lett.* **111**, 176802 (2013).

[18] Sessi , P. Reis, F., Bathon, T., Kokh, K. A., Tereshchenko, O. E. & Bode, M. Signatures of dirac fermion mediated magnetic order. *Nat. Comm.* **5**, 5349 (2014).

[19] Black-Schaffer, A. M., Balatsky, A. V. & Fransson, J. Filling of magnetic –impurity-induced gap in topological insulators by potential scattering. *Phys. Rev. B* **91**, 201411(R) (2015).

[20] Jiang, Y. *et al.* Fermi-level tuning of epitaxial $Sb_2Te_3$ thin films on graphene by regulating intrinsic defects and substrate transfer doping. *Phys. Rev. Lett.* **108**, 066809 (2012).

[21] Kachel, T., Eggenstein, F. & Follath, R. A soft x-ray plane-grating monochromator optimized for elliptical dipole radiation from modern sources. *J. Synchrotron Rad.* **22**, 1301-1305 (2015).

[22] Gambardella, P. *et al.* Ferromagnetism in one-dimensional monoatomic metal chains. *Nature* **416**, 301-304 (2002).

[23] Sessi, P. *et al.* Scattering properties of the three-dimensional topological insulator Sb2Te3: coexistence of topologically trivial and nontrivial surface states with opposite spin-momentum helicity. *Phys. Rev. B* **93**, 035110 (2016).

[24] Alpichshev, Z. *et al.* STM imaging of impurity resonances on $Bi_2Se_3$. *Phys. Rev. Lett.* **108**, 206402 (2012).

[25] Wehling, T. O., Black-Schaffer, A. M. & Balatsky, A. V. Dirac materials *Adv. Phys.* **63**, 1-76 (2014).

[26] Jiang, Y. *et al.* Landau quantization and the thickness limit of topological insulator thin films of $Sb_2Te_3$. *Phys. Rev. Lett.* **108**, 016401 (2012).

[27] Miller, D. L. *et al.* Observing the quantization of zero mass carriers in graphene. *Science* **324**, 924-927 (2009).

[28] Das Sarma, S. & Pinczuk, A. *Perspectives in quantum hall effects* (Wiley-VCH, 2004).

[29] Kokh, K. A., Makarenko, S. V., Golyashov, V. A., Shegai, O. A. & Tereshchenko, O. E. Melt growth of bulk $Bi_2Te_3$ crystals with a natural p–n junction. *CrystEngComm* **16** 581-584 (2014).

## Supplementary theoretical modeling

We start with a general low energy effective Hamiltonian of strong topological insulator (STI) surface electronic states, in the presence of impurities that have both local magnetic and potential scattering components [S1,S2],

$$\mathcal{H} = (k_x \sigma_x + k_y \sigma_y) + \sum_{\text{imp}} [J \mathbf{S}_i \cdot \boldsymbol{\sigma}(\mathbf{r}_i) + U n(\mathbf{r}_i)]$$

(S1)

Here, J and U are the impurity magnetic exchange and potential strengths, respectively. $n(r_i)$ and $\sigma(r_i)$ are the electronic number and spin densities, respectively, at the location, $r_i$, of the $i^{th}$ impurity. We have chosen units such that the Planck's constant, $\hbar$, and the Fermi (Dirac) velocity, $v_D$, are both unity.

When the impurities align ferromagnetically perpendicular to the surface (z-direction), due to RKKY interactions [S1] or other effects, it is usually assumed that an energy gap appears at the Dirac point due to the spatially averaged carrier-impurity exchange interaction. However, this analysis misses the role of impurity potential scattering, that lead to intragap resonances [S1,S2]. Taking the potential part into account, the effective Hamiltonian, including a mean field gap arising from the magnetic moments, becomes

$$\mathcal{H}_{\text{eff}} = (k_x \sigma_x + k_y \sigma_y + \Delta \sigma_z) + \sum_{\text{imp}} [J_{\text{eff}} S \sigma_z + U] n(\mathbf{r}_i)$$

(S2)

The mean field Dirac mass,

$$\Delta \simeq J \sum_{\text{imp}} \langle n(\mathbf{r}_i)(\mathbf{S}_i)_z \rangle ,$$

(S3)

is half the energy gap of the (now massive) surface Dirac states, calculated using an appropriate averaging procedure over the impurity-carrier exchange interactions. $J_{\text{eff}}$ is the residual effective magnetic impurity strength of the impurity acting on these *massive* Dirac fermions.

Ignoring the chemical potential shift, a mean field effect arising from potential scattering, as well as residual local impurity effects, these massive Dirac fermions possess a fully gapped spectrum:

$$E(k) = \pm\sqrt{\Delta^2 + k^2}$$

(S4)

However, as we show now, completely ignoring the local residual effects of potential scattering is a mistake. In general, no magnetic impurity should have U = 0. The inclusion of U leads to the formation of impurity resonances in gapless Dirac Materials [S1,S3]. Numerical analysis [S2] shows that in the presence of both magnetic and potential scattering, the energy spectrum contains both a gapped feature due to magnetism, as well as the aforementioned impurity resonances due to potential scattering. As a result, the spectrum is not fully gapped, a qualitative difference from the mean field result.

Using the effective Hamiltonian, Eq. (S2), in the presence of a single impurity, we can exactly calculate the single particle Green's function using the T-matrix technique [S1,S4]. We find analytically that the T-matrix is singular, i.e., there exists a bound state, when the energy, E, is a solution to the equation,

$$g(E)(J_{\text{eff}}S \pm U)(\Delta \pm E) = 1,$$
(S5)

while also being inside the gap, $-\Delta<E<\Delta$. $g(E)$ is the on-site Green's function, which for $-\Delta<E<\Delta$ is the following

$$g(E) = -\frac{1}{4\pi} \ln\left(1 + \frac{\Lambda^2}{\Delta^2 - E^2}\right) < 0.$$
(S6)

$\Lambda$ is the high energy cutoff. These equations always possess a solution when U is larger than $J_{\text{eff}}S$ in magnitude, and so there is a inter-gap bound state for a range of impurity parameters.

Including appropriate spectral broadening of the Dirac states, as is observed experimentally, we present the calculated spectra for a range of values of $U/J_{\text{eff}}S$, in Figs S1 and S2. For these calculations we have used units in which the Planck's constant, $\hbar$, the Fermi velocity, $v_D$, as well as the mean field energy gap, $\Delta$, are unity. We have used the values $\Lambda=50$, $J_{\text{eff}}S=10$ and the spectral broadening, $\Gamma=0.3$. Impurities are assumed to be located at distances r = 0.6, 1.4, 1.5, 1.6, 1.8, 1 and 3; these locations are chosen randomly to yield LDOS that resemble Fig 2d in the main text and are consistent with the neighborhood of the location where that data was measured. Multiple impurity scattering has been ignored. As U becomes large in magnitude compared to $J_{\text{eff}}S$, a bound state peak appears inside the energy gap.

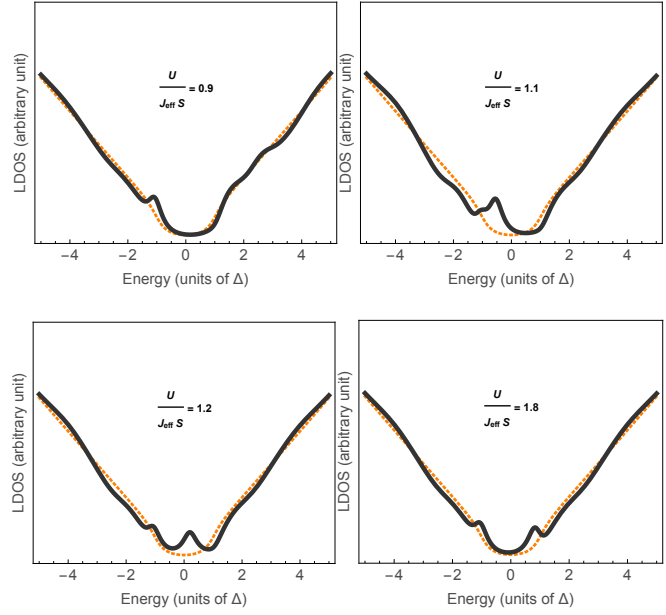

Figure S1: The LDOS at a location with many nearby impurities, for various positive ratios of the scalar potential impurity strength, U, and the residual impurity exchange strength, $J_{eff}S$. $\Delta$ is the mean field mass of the Dirac gas, which arises due to the averaged exchange field seen by the STI surface states in the presence of ferromagnetically aligned impurity spins which point perpendicular to the surface. The dashed curve is the LDOS in the absence of the local impurity perturbation, which exhibits an energy gap between $\pm\Delta$. We see that when the ratio $U/(J_{eff}S)$ is greater than a certain value (equal to 1 for no energy broadening), bound states appear inside the energy gap. Thus, experimental measurements of the LDOS will may not observe a clear spectral gap, even though the mobility gap is nonzero, because the inter-gap bound states do not contribute to transport.

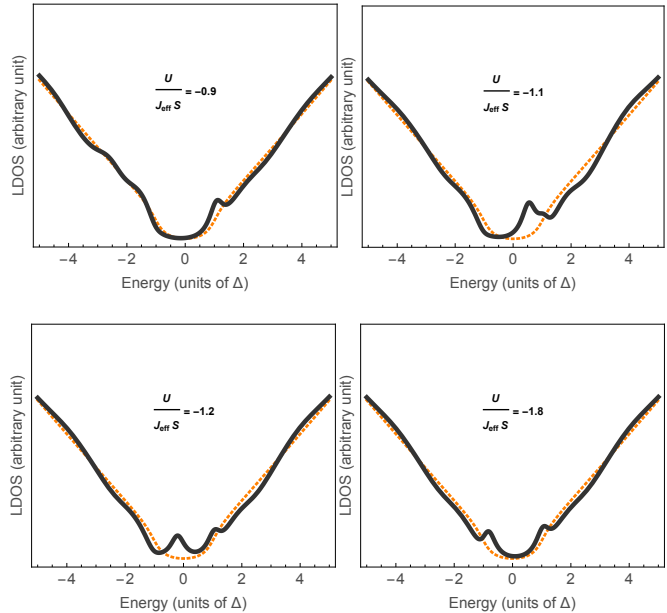

Figure S2: The same as Fig S1, but with negative values of the ration $U/(J_{eff}S)$.

Our calculations thus directly prove that for a range of reasonable parameters, (when the magnitude of U is larger than that of $J_{eff}S$, for $\Gamma=0$), while the spectrum of *extended* states may possess an energy gap (the mobility gap), the LDOS need not exhibit an energy gap. This reflects the competing trends between gap opening due to magnetic coupling, and gap filling due to localized states seeded from the scalar potential-induced spectral resonances in the gapless case

**Supplementary figure 3**

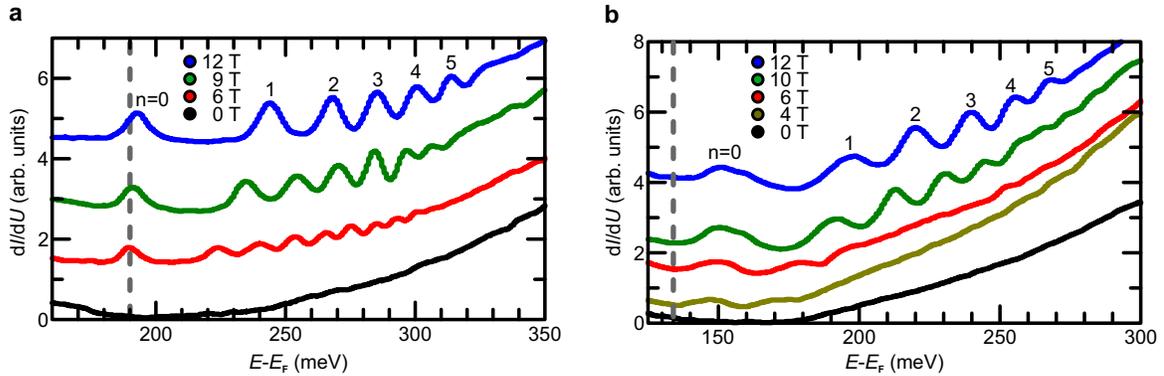

Figure S3: Scanning tunneling spectroscopy data obtained on a pristine and b V-doped $Sb_2Te_3$ at different magnetic fields applied perpendicular to the sample surface. Spectra at different magnetic fields have been vertically shifted for clarity. For sufficiently large fields, a well-defined sequence of peaks emerge form the background, signaling the condensation of the two-dimensional topological electron gas into Landau orbits. The index n identifies the Landau level order. The dashed gray line marks the position of the Dirac point as expected for linearly dispersing massless particles. In V-doped samples, its shift with respect to the position of the first peak (n=0) unequivocally signals the emergence of a mass term in the Dirac spectrum.

**References:**

[S1] R. R. Biswas and A. V. Balatsky Impurity-induced states on the surface of three-dimensional topological insulators Phys. Rev. B 81, 233405 (2010).

[S2] A. M. Black-Schaffer et al. Filling of magnetic –impurity-induced gap in topological insulators by potential scattering Phys. Rev. B 91, 201411(R) (2015).

[S3] T. O. Wehling, A. M. Black-Schaffer and A. V. Balatsky, Dirac Materials Adv. Phys. 76, 1 (2014).

[S4] S. Doniach and E. H. Sondheimer, Green's Functions for Solid State Physicists, Imperial College Press, London (1998).